\documentclass[a4paper,fleqn,usenatbib]{mnras}

\usepackage{newtxtext,newtxmath}

\usepackage[T1]{fontenc}
\usepackage{ae,aecompl}

\usepackage{graphicx}	
\usepackage{amsmath, bm}	
\usepackage{amssymb}	

\title[Whistler heat flux saturation in the solar wind]{Quasilinear Approach of the Whistler Heat-Flux Instability in the Solar Wind}

\author[S.M.Shaaban et al.]{
S. M. Shaaban,$^{1,2}$\thanks{E-mail: shaaban.mohammed@kuleuven.be}
M. Lazar,$^{1,3}$
P.H.Yoon,$^{4,5,6}$
S. Poedts,$^{1}$
R. A. L{\'o}pez$^{1}$
\\
 $^{1}$Centre for Mathematical Plasma Astrophysics, KU Leuven, Celestijnenlaan 200B, B-3001 Leuven, Belgium.\\
$^{2}$Theoretical Physics Research Group, Physics Department, Faculty of Science, Mansoura University, 35516, Mansoura, Egypt.\\
$^{3}$Institut f\"ur Theoretische Physik, Lehrstuhl IV: Weltraum- und Astrophysik, Ruhr-Universit\"at Bochum, D-44780 Bochum, Germany.\\
$^{4}$Institute for Physical Science and Technology, University of Maryland, College Park, MD 20742, USA.\\
$^{5}$Korea Astronomy and Space Science Institute, Daejeon 34055, Republic of Korea.\\
$^{6}$School of Space Research, Kyung Hee University, Yongin, Gyeonggi 17104, Republic of Korea.
}

\date{Accepted XXX. Received YYY; in original form ZZZ}

\pubyear{2019}

\begin{document}
\label{firstpage}
\pagerange{\pageref{firstpage}--\pageref{lastpage}}
\maketitle

\begin{abstract}
The hot beaming (or strahl) electrons responsible for the main electron heat-flux in the 
solar wind are believed to be self-regulated by the electromagnetic beaming instabilities, 
also known as the heat-flux instabilities. Here we report the first quasi-linear theoretical 
approach of the whistler unstable branch able to characterize the long-term saturation of 
the instability as well as the relaxation of the electron velocity distributions. 
The instability saturation is not solely determined by the drift velocities, which 
undergo only a minor relaxation, but mainly from a concurrent interaction of electrons 
with whistlers that induces (opposite) temperature anisotropies of the core and beam
populations and reduces the effective anisotropy. These results might be able to (i) explain 
the low intensity of the whistler heat-flux fluctuations in the solar wind (although other 
explanations remain possible and need further investigation), and (ii) confirm a reduced 
effectiveness of these fluctuations in the relaxation and isotropization of the electron 
strahl and in the regulation of the electron heat-flux.

\end{abstract}

\begin{keywords}
instabilities -- solar wind -- methods: numerical
\end{keywords}

%
\section{Introduction}
%
Guided by the interplanetary magnetic field the electron beaming or \textit{strahl} population carries 
the main electron heat flux in the solar wind \citep{Feldman1975, Lin1998, Pierrard2001}. At large 
distances from the Sun in-situ measurements reveal a significant inhibition of the electron heat flux 
\citep{Gary1999, Scime1994, Bale2013} below the Spitzer--H\"arm predictions \citep{Spitzer1953}. 
Particle-particle collisions are rare and inefficient at large heliocentric distances, but the heat-flux
can be regulated by the self-generated instabilities, through the wave-particle interactions.   
The observations support this hypothesis, showing evidences of enhanced electromagnetic fluctuations
usually associated with the heat-flux instabilities \citep{Scime1994, Gary1999, Pagel2007, Bale2013, 
Lacombe2014}. Numerical simulations confirm a potential role of these instabilities in the regulation
of electron heat-flux in the solar wind \citep{Vocks2005, Saito2007, Roberg2018a}.

Heat-flux instabilities can manifest either as a whistler growing mode or as a firehose-like instability
\citep{Gary1985, Saeed2017a, Saeed2017b, Shaaban2018HFA, Shaaban2018HF}, but such a distinction, although 
useful, is not always taken into account in specific studies of these instabilities. Here we focus 
on the whistler heat-flux instability, predicted by the linear theory for less energetic beams, i.e.,
with drifting velocity lower than thermal speed  \citep{Gary1985, Gary1999, Saeed2017a, Shaaban2018HF, 
Tong2018}, and often invoked to explain the enhanced fluctuations observed in the slow and moderate 
winds, e.g., with $v_{SW} < 500$ km/s \citep{Pagel2007, Lacombe2014, Stansby2016, Tong2019}. These 
fluctuations may pitch-angle scatter the strahl, which becomes broader as electron energy 
increases and reduces in intensity with heliocentric distance \citep{Maksimovic2005, Vocks2005, 
Pagel2007}, explaining thus the inhibition of the electron heat flux in the solar wind \citep{Gary1985, 
Tong2018, Tong2019}. 

Whistler waves (also known as electromagnetic electron cyclotron modes) are observed propagating 
at small angles along the interplanetary magnetic field with a right-handed circular polarization and 
frequency between ion and electron gyrofrequencies, i.e.\ $\Omega_p < \omega < |\Omega_e|$ 
\citep{Wilson2013, Lacombe2014}. Local sources of whistlers are multiple, either the heat-flux 
instability driven by counter-beaming electrons \citep{Feldman1973, Gary1993, Shaaban2018HF}, or the
cyclotron instability driven by electrons with anisotropic temperatures, i.e., $T_\perp> T_\parallel$, 
where $\parallel, \perp$ denote directions with respect to the magnetic field \citep{Gary1996, 
Lazar2018}, or even the interplay of these two instabilities \citep{Shaaban2018HFA}. 
Both sources of free energy may indeed co-exist in space plasmas \citep{Stverak2008, Vinas2010, Lazar2017},
and recent studies \citep{Saeed2017b, Shaaban2018HF, Shaaban2018HFA} have unveiled new 
regimes of whistler heat-flux instability in an attempt to provide an extended linear description 
in the solar wind conditions.

In the present paper we provide valuable physical insights from a quasilinear (QL) approach of the 
whistler heat-flux instability, which enable decoding of the main mechanisms leading to the 
saturation of the growing fluctuations and the relaxation of the electron counter-beaming distribution. 
As already mentioned, this instability involves less energetic beams and, implicitly, low-level
growth-rates and fluctuations which are not easily captured in the simulations. In this context,
a quasilinear approach can offer unique methods to investigate the long-term evolution of 
this instability and its actions back on the electron velocity distribution.
Quasilinear approaches have been successfully employed in studies of both the whistler and firehose 
instabilities driven by the temperature anisotropy, and the results showed agreements with the linear theory predictions, particle-in-cell simulations, and the observational limits of the electron temperature 
anisotropy \citep{Yoon2012, Seough2014, Seough2015, Yoon2017, Lazar2018, Shaaban2019}. 

The manuscript is structured as follows: In section~\ref{Sec:2} we introduce the particle velocity 
distribution functions (VDFs), with a focus on the counter-drifting (dual) core-beam model for 
the electrons. This model describes the counter-moving of the core and beam populations 
in terms of their drifting velocities, i.e.\ $U_c$ and $U_b$, respectively, which act as a source of 
free energy triggering different instabilities. Both linear and QL theoretical formalisms for 
dispersion and stability are  are described in section~\ref{Sec:3}. In section~\ref{Sec:4} we present 
an extended analysis of the unstable whistler heat-flux solutions for three cases with potential implications
 in the regulation of the electron strahl and the electron heat flux in the solar wind. The results obtained in
the present work are summarized in section~\ref{Sec:5}.

\section{Counter-drifting solar wind electron distribution}\label{Sec:2}
%
Solar wind in-situ measurements, e.g., from various spacecraft missions, e.g., \textit{Helios 1}, 
\textit{Cluster II}, \textit{ Ulysses}, or \textit{Wind}, reveal electron distributions 
with a dual structure combining two counter-drifting components in a frame 
fixed to protons, namely, a thermal and dense core and a strahl population 
streaming along the magnetic field \citep{Gary1999, Maksimovic2005, Tong2019}. 
\begin{equation}\label{e1}
f_e\left(v_\parallel,v_\perp \right)=\dfrac{n_c}{n_0}~f_c\left(v_\parallel,v_\perp 
\right)+ \dfrac{n_b}{n_0} ~f_b\left(v_\parallel,v_\perp \right).  
\end{equation}
where $n_c$ and $n_b$ are the core and beam number density, respectively, and 
$n_0$ is the total number density of electrons. Quasilinear approach is not 
straightforward, but for the sake of simplicity, here we assume both the core 
(subscript $c$) and beam (subscript $b$) components well described by drifting
bi-Maxwellian models \citep{Shaaban2018HFA}
\begin{align}
 \label{e2}
f_{a}\left( v_{\parallel },v_{\perp }\right) =&\frac{1}{\pi
^{3/2} \alpha_{\perp a }^{2} ~ \alpha_{\parallel a }}\exp \left(
-\frac{v_{\perp
}^{2}}{\alpha_{\perp a}^{2}}-\frac{\left(v_{\parallel }- U_a\right)^{2}} {\alpha_{\parallel a}^{2}}\right)   
\end{align}
where thermal velocities $\alpha_{ \parallel, \perp, a} \equiv\alpha_{\parallel, 
\perp, a}(t)$ are defined in terms of the corresponding kinetic temperature components, which may evolve in time ($t$)
\begin{subequations}\label{e3}
\begin{align}
T_{\parallel, a}&=\frac{m_e}{k_B}\int d\textbf{v}~(v_\parallel-U_a)^2 
~f_a(v_\parallel, v_\perp)=\frac{m_e ~\alpha_{\parallel, a}^2}{2 k_B},\\
T_{\perp, a}&=\frac{m_e}{2 k_B}\int d\textbf{v}~ v_\perp^2
~f_a(v_\parallel, v_\perp)=\frac{m_e ~\alpha_{\perp, a}^2}{2 k_B},
\end{align}
\end{subequations}
$U_a$ is drifting velocity, either for the core (subscript "$a=c$") or 
beam (subscript "$a=b$"), along the background magnetic field.
We perform our analysis in a quasi-neutral electron-proton plasma 
$n_e=n_c+n_b\approx n_p$, with zero net current, i.e.\ $n_c~U_c+n_b~U_b=0$.

%
\section{Quasilinear instability approach}\label{Sec:3}
%
For a collisionless and homogeneous plasma the linear dispersion relation derived
from kinetic theory for the right-handed (RH) circularized polarized electromagnetic
modes propagating in directions
parallel to the stationary magnetic field (${\bf k} \times {\bf B}_0 = 0$)
reads \citep{Shaaban2018HFA}
\begin{align} \label{e4}
\tilde{k}^2=&(1-\delta)~\mu~ \left[\Lambda_c+\frac{(\Lambda_c+1)~(\tilde{\omega}-\tilde{k}~u_c) -\mu~ \Lambda_c}{\tilde{k} \sqrt{\mu~\beta_c}}\right.\nonumber\\
&\left.\times Z_c\left(\frac{\tilde{\omega}-\mu-\tilde{k}~u_c}{\tilde{k} \sqrt{\mu~\beta_c}}\right)\right]+\frac{\tilde{\omega}}{\tilde{k}~\sqrt{\beta_p}}Z_p\left(\frac{\tilde{\omega}+1}{\tilde{k} \sqrt{\beta_p}}\right)\nonumber\\
&+\delta~\mu \left[\Lambda_b+\frac{(\Lambda_b+1)~(\tilde{\omega}-\tilde{k}~u_b) -\mu~ \Lambda_b}{\tilde{k} \sqrt{\mu~\beta_b}}\right.\nonumber\\
&\left.\times Z_b\left(\frac{\tilde{\omega}-\mu-\tilde{k}~u_b}{\tilde{k} \sqrt{\mu~\beta_b}}\right)\right]
\end{align}
where $\tilde{k}=kc/\omega_{p,p}$ is the normalization used for wave-number $k$, 
$c$ is the speed of light, $\omega_{p, p}~=~\sqrt{4\pi n_0 e^2/m_p}$ is the plasma 
frequency of protons, $\tilde{\omega}=\omega/\Omega_p$ is the normalization for wave 
frequency, $\Omega_p$ is the non-relativistic gyro-frequency of protons, 
$\mu=m_p/m_e$ is the proton--electron mass contrast, $\Lambda_a=~T_{\perp,a}/
T_{\parallel,a}-~1\equiv A_a-1$ and $\beta_{\parallel,\perp , a}=8\pi n_0 k_B 
T_{\parallel \perp , a}/B_0^2$ are, respectively, the temperature anisotropy, and 
plasma beta parameters for protons (subscript "$a=p$"), electron core (subscript 
"$a=c$"), and electron beam (subscript "$a=b$") populations, $\delta=n_b/n_0$, 
$1-\delta=n_c/n_0$ are the beam and core relative densities, respectively,  
$u_a=U_a/v_A$ are normalized drifting velocities, $v_A=\sqrt{B_0^2/4\pi n_p m_p}$ is the proton Alfv{\'e}n speed, and
\begin{equation}  \label{e5}
Z_{a}\left( \xi _{a}^{\pm }\right) =\frac{1}{\sqrt{\pi}}\int_{-\infty
}^{\infty }\frac{\exp \left( -x^{2}\right) }{x-\xi _{a}^{\pm }}dt,\
\ \Im \left( \xi _{a}^{\pm }\right) >0, 
\end{equation}
are plasma dispersion functions \citep{Fried1961}. 

In the quasilinear formalism one must solve for both particle and wave kinetic 
equations. For parallel propagation of electromagnetic waves, the particle kinetic equation 
in the diffusion approximation describes the time evolution of the velocity distributions 
as follows 
\begin{align} \label{e6}
\frac{\partial f_a}{\partial t}&=\frac{i e^2}{4m_a^2 c^2~ v_\perp}\int_{-\infty}^{\infty} 
\frac{dk}{k}\left[ \left(\omega^\ast-k v_\parallel\right)\frac{\partial}{\partial v_\perp}+ 
k v_\perp\frac{\partial}{\partial v_\parallel}\right]\nonumber\\
&\times~\frac{ v_\perp \delta B^2(k, \omega)}{\omega-kv_\parallel-\Omega_a}\left[ 
\left(\omega-k v_\parallel\right)\frac{\partial f_a}{\partial v_\perp}+ k v_\perp
\frac{\partial f_a}{\partial v_\parallel}\right]
\end{align}
where $\delta B^2(k)$ is the energy density of the fluctuations. The wave equation is 
given by 
\begin{equation} \label{e7}
\frac{\partial~\delta B^2(k)}{\partial t}=2 \gamma_k \delta B^2(k),
\end{equation}
with growth rate $\gamma_k$ of the unstable whistler solutions obtained from Eq.~\eqref{e4}. 
Dynamical equations for the macroscopic moments of the veocity distribution,
such that the drift velocities $U_a$ of core (subscript "$a=c$") and beam (subscript "$a=b$"), 
and their temperature components $T_{\perp, \parallel, a}$ are derived from Eq.~\eqref{e6} 
as follows
\begin{subequations}\label{e8}
\begin{align}
\frac{dU_{c}}{dt}&=\frac{e^2 m_e^{-2}}{2~ c^2}
\int_{-\infty}^{\infty}\frac{dk}{k}\langle~ \delta B^2(k)~\rangle\text{Im} \left\lbrace\eta_c~ 
Z_{c}\left(\zeta_c\right)\right\rbrace\\
\frac{dU_{b}}{dt}&=\frac{e^2 m_e^{-2}}{2~ c^2}
\int_{-\infty}^{\infty}\frac{dk}{k}\langle~ \delta B^2(k)~\rangle \text{Im} \left\lbrace \eta_h~ 
Z_{b}\left(\zeta_b\right)\right\rbrace\\
\frac{dT_{\perp a}}{dt}&=-\frac{e^2}{2m_e c^2}
\int_{-\infty}^{\infty}\frac{dk}{k^2}\langle~ \delta B^2(k)~\rangle \nonumber\\
&\times\left\lbrace\left(2 \Lambda_c+1\right)\gamma_k+\text{Im} 
\left(2i\gamma-\Omega_e\right)~\eta_a Z_a\left(\zeta_a\right)\right\rbrace\\
\frac{dT_{\parallel a}}{dt}&=\frac{e^2}{m_e c^2}
\int_{-\infty}^{\infty}\frac{dk}{k^2}\langle~ \delta B^2(k)~\rangle\nonumber\\
&\times \left\lbrace 2\left(\Lambda_a+1\right)\gamma_k+\text{Im}~k~\alpha_{\parallel a}~
\zeta_a ~ \eta_a Z_a\left(\zeta_a\right)\right\rbrace
\end{align}
\end{subequations}
with
\begin{align*}
&\eta_c=\left[\left(\Lambda_c+1\right)\left(\omega-k~U_c\right) -  \Omega_e~\Lambda_c\right]/\left(k ~\alpha_{\parallel c}\right),\\
&\eta_b=\left[\left(\Lambda_b+1\right)\left(\omega-k~U_b\right) -  
\Omega_e~\Lambda_b\right]/\left(k ~\alpha_{\parallel b}\right),\\
&\zeta_c=\frac{\omega-\Omega_e -k~U_c}{k~\alpha_{\parallel c}} ~~ \& ~~ 
\zeta_b=\frac{\omega-\Omega_e -k~U_b}{k~\alpha_{\parallel b}}.
\end{align*}
As above in Eq.~\eqref{e4}, we can use normalized quantities
\begin{align*}
&\tilde{\eta}_c=\left[\left(\Lambda_c+1\right)\left(\tilde{\omega}-\tilde{k}~u_c\right) -  \mu~\Lambda_c\right]/\left(\tilde{k}~ \sqrt{\mu~\beta_{\parallel c}}\right),\\
&\tilde{\eta}_b=\left[\left(\Lambda_b+1\right)\left(\tilde{\omega}-\tilde{k}~u_b\right) -  
\mu~\Lambda_b\right]/\left(\tilde{k}~ \sqrt{\mu~\beta_{\parallel b}}\right),\\
&\tilde{\zeta}_c=\frac{\tilde{\omega}-\mu -\tilde{k}~u_c}{\tilde{k}~\sqrt{\mu~\beta_{\parallel c}}}, ~~ 
\tilde{\zeta}_b=\frac{\tilde{\omega}-\mu -\tilde{k}~u_b}{\tilde{k}~\sqrt{\mu~\beta_{\parallel b}}}\\
&u_b=U_b/v_A,~~ u_c=U_c/v_A, ~~ \tau=\Omega_p~t
\end{align*}
and $W(\tilde{k})=\delta B^2(\tilde{k})/B_0^2$ (for the wave energy density), to find
\begin{subequations}\label{e9}
\begin{align}
\frac{du_{c}}{d\tau}&=\mu^2
\int_{\infty}^{\infty}\frac{d\tilde{k}}{\tilde{k}} W\left(\tilde{k}\right)\text{Im} 
\left\lbrace \tilde{\eta}_c Z_{c}\left(\tilde{\zeta}_c\right)\right\rbrace\\
\frac{du_{b}}{d\tau}&=\mu^2
\int_{-\infty}^{\infty}\frac{d\tilde{k}}{\tilde{k}} W\left(\tilde{k}\right)\text{Im} 
\left\lbrace  \tilde{\eta}_b Z_{h}\left(\tilde{\zeta}_b\right)\right\rbrace\\
\frac{d\beta_{\perp a}}{d\tau}&=-2\mu\int_{-\infty}^{\infty}\frac{d\tilde{k}}{\tilde{k}^2}	~W\left(\tilde{k}\right)\nonumber \\
&\times\left\lbrace\left(2 \Lambda_c+1\right)\tilde{\gamma}_k+\text{Im} \left(2i\tilde{\gamma}-\mu\right)\tilde{\eta}_a Z_a\left(\tilde{\zeta}_a\right)\right\rbrace\\
\frac{d\beta_{\parallel a}}{d\tau}&=4\mu\int_{-\infty}^{\infty}\frac{d\tilde{k}}{\tilde{k}^2}	~W\left(\tilde{k}\right) \nonumber \\
&\times\left\lbrace 2 \left(\Lambda_c+1\right)\tilde{\gamma}_k+\text{Im}~\tilde{k}\sqrt{\mu~\beta_{\parallel a}} ~\tilde{\zeta}_a \tilde{\eta}_a Z_a\left(\tilde{\zeta}_a\right)\right\rbrace
\end{align}
\end{subequations}
and 
\begin{eqnarray}\label{e10}
\frac{\partial~W(\tilde{k})}{\partial \tau}=2~\tilde{\gamma}~ W(\tilde{k}).
\end{eqnarray}

If both the core and beam populations are assumed drifting bi-Maxwellian the heat flux is 
given by \citep{Gary1994}
\begin{align}\label{e11}
&q_e=\dfrac{m_e}{2}\sum_{a=c,b} n_a U_a\left[\left(3+2\dfrac{T_{\perp a}}{T_{\parallel a}}\right) 
\alpha_{\parallel a}^2+U_a^2\right]
\end{align}
or normalized %
\begin{align}\label{e12}
&\dfrac{q}{q_{max}}=\left(1-\delta\right) \dfrac{|u_c|}{\sqrt{\mu \beta_{\parallel c}}}
\left[\left(\dfrac{\beta_{\parallel b}}{\beta_{\parallel c}}-1\right) +\dfrac{2}{3}\left(\dfrac 
{\beta_{\perp b}}{\beta_{\perp c}}-1\right)\dfrac{\beta_{\perp c}}{\beta_{\parallel c}}\right],
\end{align}
using $q_{max}=3 n_0 T_{\parallel c}\alpha_{\parallel c}/2$.
For isotropic temperatures $\beta_{\perp} = \beta_{\parallel}$, and Eq.(\ref{e12}) reduces to
\begin{align}\label{e13}
&\dfrac{q}{q_{max}}=\dfrac{5}{3} ~\left(1-\delta\right)~ \dfrac{|u_c|}{\sqrt{\mu 
~\beta_{\parallel c}}}\left(\dfrac{\beta_{\parallel b}}{\beta_{\parallel c}}-1\right),
\end{align}
%

%
\section{Numerical results} \label{Sec:4}
In this section we discuss the numerical results from the linear and QL analysis of 
the unstable whistler heat flux mode triggered by the core-beam counterstreaming 
electrons. Details are presented for three distinct cases corresponding to conditions typically 
encountered in the solar wind \citep{Maksimovic2005, Tong2018}, with the following 
plasma parameters:
\begin{enumerate}
\item{Case 1.}
 \begin{eqnarray}\label{e14}
&&U_{b}(0)=40 ~v_A , \text{and}~ U_{c}(0)=-2.1 ~v_A \nonumber\\
&&\beta_c(0)=1,~ 2,~ 3, \text{and}~T_b(0)=6~T_c(0),\nonumber\\
&&A_{c,b}(0, \tau_{max})=\frac{\beta_{\perp, c,b}(0, \tau_{max})}{\beta_{\parallel, c, 
b}(0, \tau_{max})}=1.0,
\end{eqnarray}

\item{Case 2.}
 \begin{eqnarray}\label{e15}
&&U_{b}(0)/v_A=60,~ 40, ~20 , \nonumber\\
&& U_{c}(0)/v_A=-3.1,~ -2.1,~ -1.0, \nonumber\\
&&\beta_c(0)= 2, \text{and}~T_b(0)=6~T_c(0),\nonumber\\
&&A_{c,b}(0, \tau_{max})=\frac{\beta_{\perp, c,b}(0, \tau_{max})}{\beta_{\parallel, c, 
b}(0, \tau_{max})}=1.0,
\end{eqnarray}

\item{Case 3. }
 \begin{eqnarray}\label{e16}
&&U_{b}(0)=40 ~v_A , \text{and}~ U_{c}(0)=-2.1 ~v_A \nonumber\\
&&\beta_c(0)=3, \text{and}~T_b(0)=6~T_c(0),\nonumber\\
&&A_{c,b}(0)=\frac{\beta_{\perp, c,b}(0)}{\beta_{\parallel, c, b}(0)}=1.0.
\end{eqnarray}
\end{enumerate}
Other plasma parameters used in our numerical computations are  $\delta=0.05$, $W(k)=5\times10^{-6}$
$\omega_{p,e}/|\Omega_e|=100$ and $v_A=2\times 10^{-4}~c$.

\begin{figure}
\centering
\includegraphics[scale=0.45, trim={3.05cm 2.7cm 2.5cm 2.7cm}, clip]{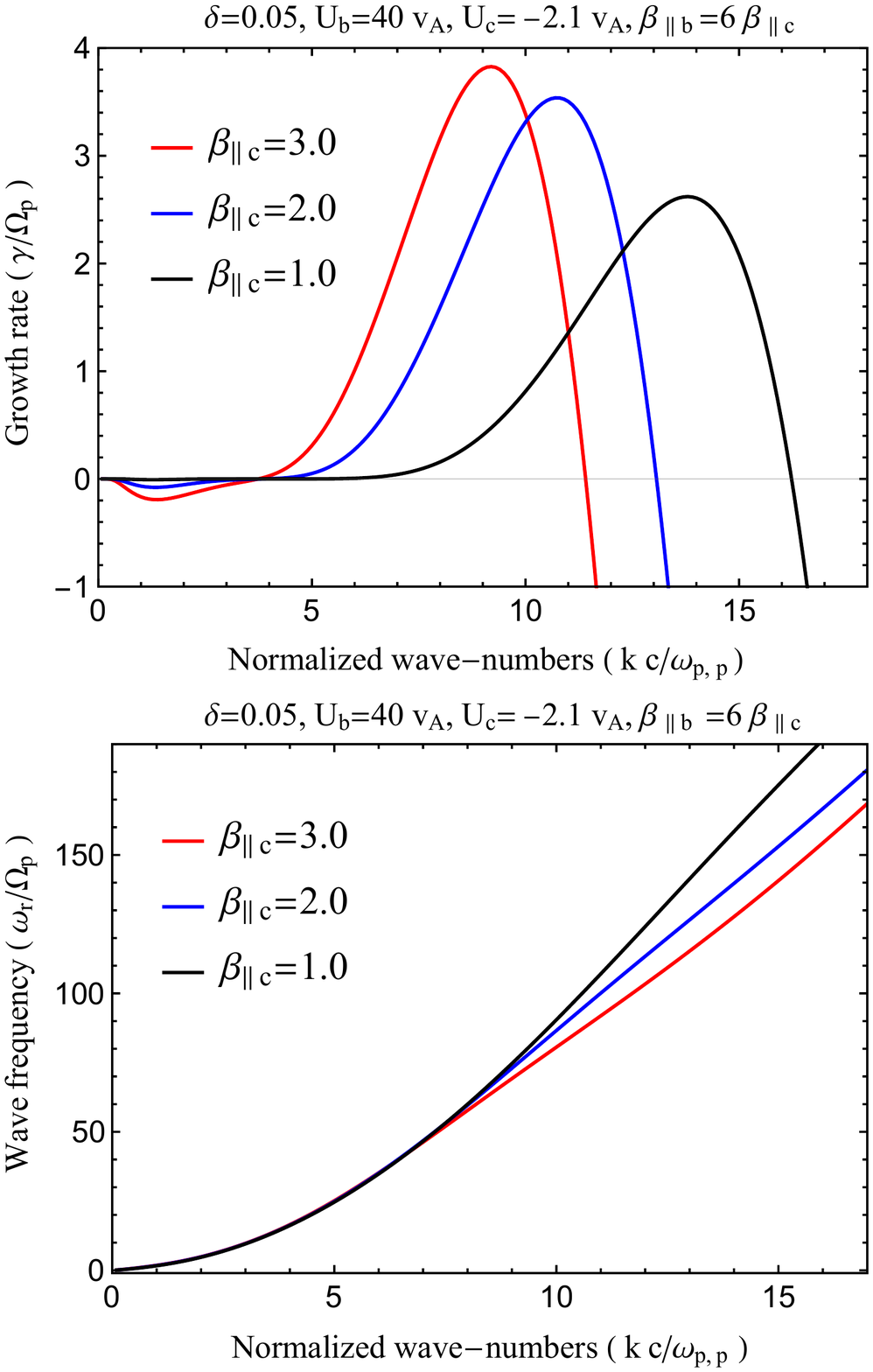}
\caption{Wave-number dispersion of whistler heat flux growth rates (top) and wave-frequencies (bottom)
and their variation with plasma beta.}\label{f1}
\end{figure}
\begin{figure}
\centering
\includegraphics[scale=0.45, trim={2.9cm 2.5cm 2.5cm 2.7cm}, clip]{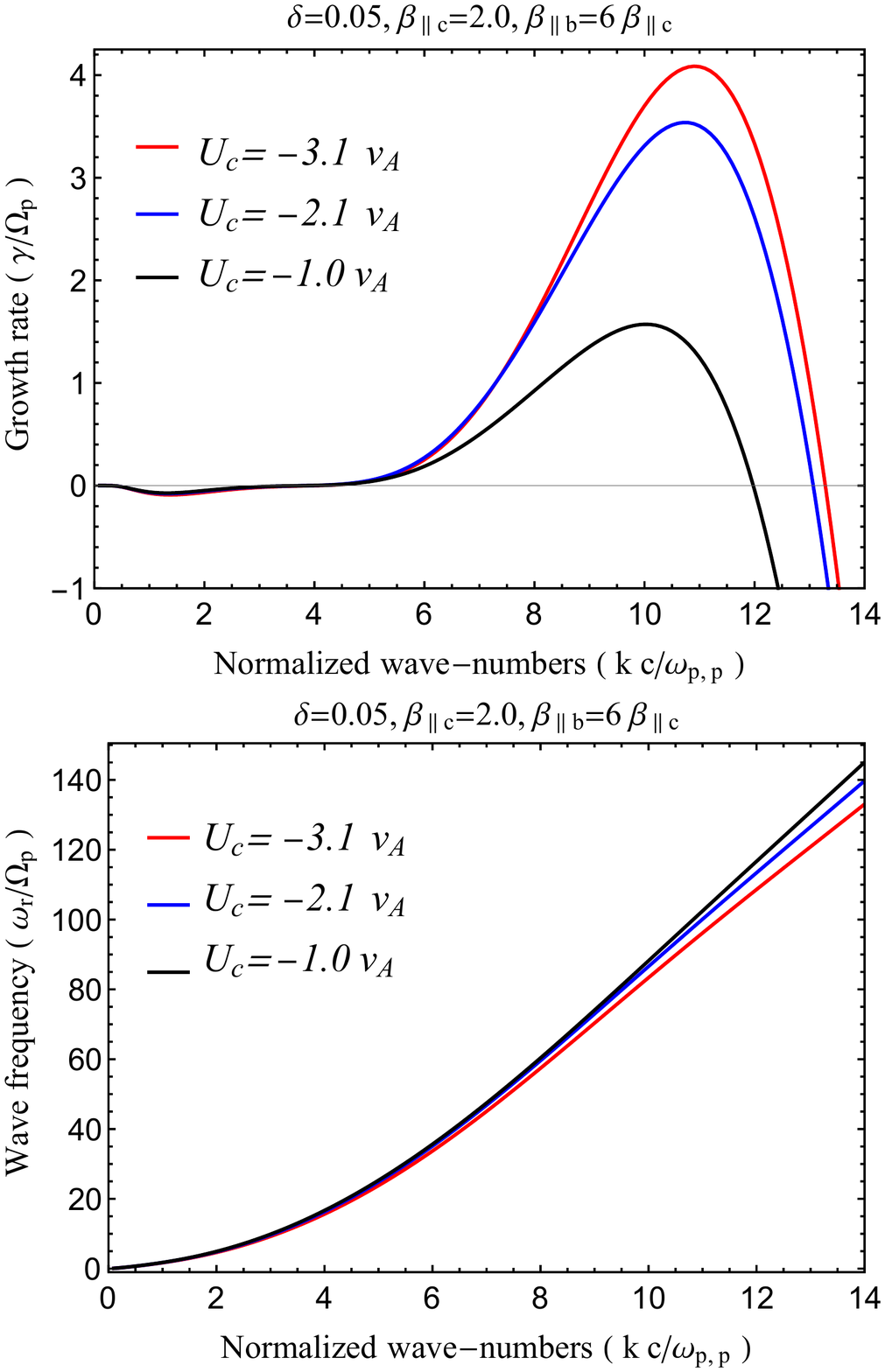}
\caption{Wave-number dispersion of whistler heat flux growth rates (top) and wave-frequencies (bottom)
and their variation with core drift velocity.}\label{f2}
\end{figure}
%

\subsection{Linear Analysis}\label{Sec:4.1}
%
\begin{figure*}
\centering
\includegraphics[scale=1.05, trim={2.6cm 14.1cm 2.cm 2.5cm}, clip]{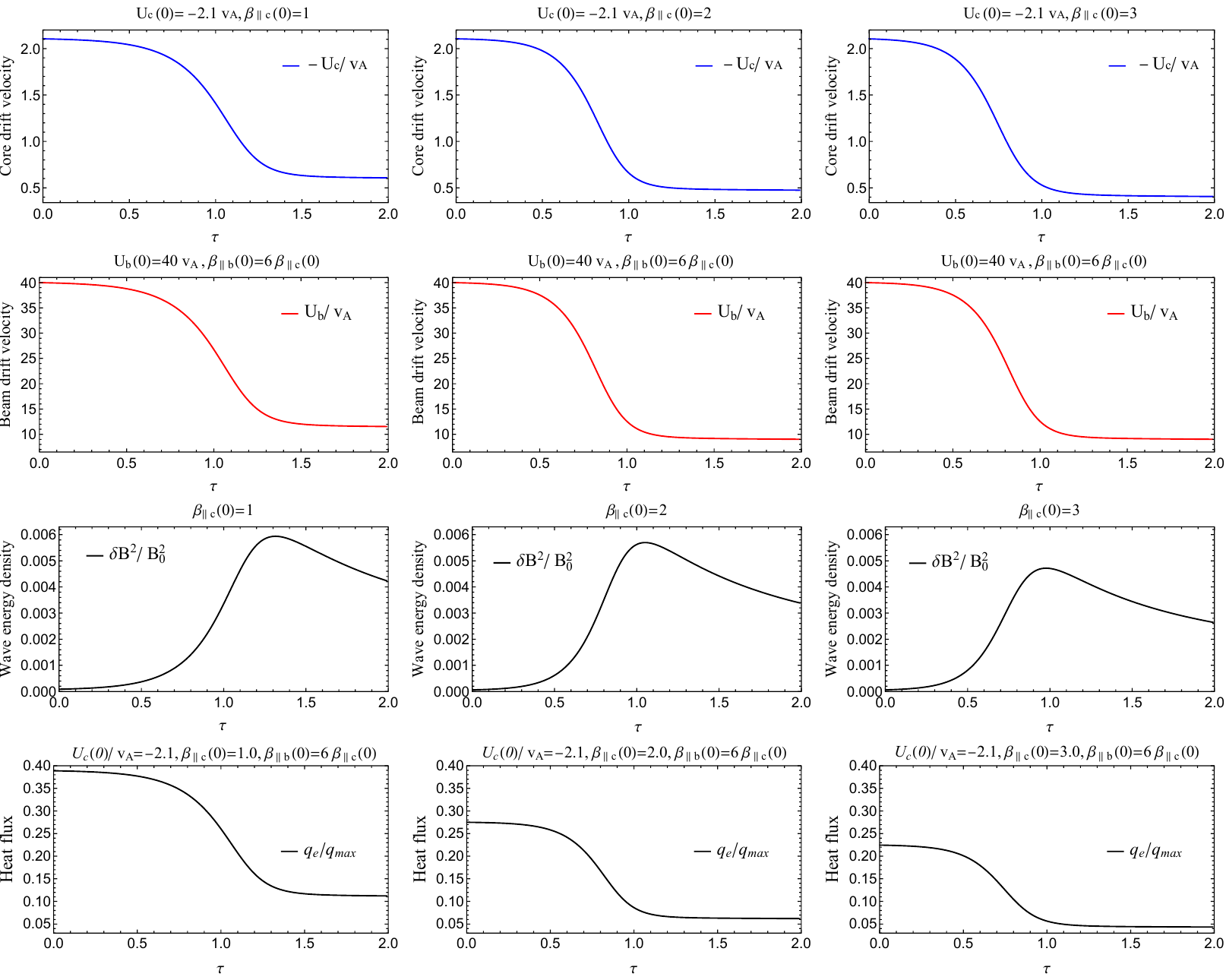}
\caption{Case~1: QL time evolution of the drift velocities for 
core ($U_c$) and beam electrons ($U_b$), the wave energy density ($\delta B(t)/B_0^2$), and the
heat flux ($q_e/q_{max}$) for different initial conditions: $\beta_c(0)=1$ (left), $2$ (middle), 
$3$ (right).} \label{f3}
\end{figure*}
For a linear analysis we solve numerically the wavenumber dispersion relation (\ref{e5}). 
In Figure~\ref{f1} we study the 
effect of the core plasma beta $\beta_{\parallel c}=1, 2, 3$ on the dispersive 
characteristics of the WHF instability, i.e., growth rate (top panel) and wave 
frequency (bottom panel), using the plasma parameters in case 1. The instability 
growth rate increases as the core plasma beta increases, but the range of the 
unstable wavenumbers decreases and the WHF instability becomes more effective at lower 
wavenumbers. The corresponding wave frequencies are decreasing with increasing 
the core plasma beta. For case 2.\ the unstable WHF solutions are displayed in Figure~\ref{f2} enabling us to compare the instability growth rates (top panel) and wave 
frequencies (bottom panel) for different core drift velocities $u_c=-3.1,~-2.1,~
-1.0$ (implying different velocities for the beam component $u_b=60,~40,~20$). 
The growth rates are increasing with increasing the core drift velocity, while 
the corresponding wave frequencies are slightly decreasing. However, for 
$\beta_{\parallel c}=2.0$ and $u_c=-3.1$ the WHF achieves its maximum growth 
rate and any further increase in the drift velocity inhibits the growth
 rate. \cite{Shaaban2018HF, Shaaban2018HFA} have described in detail this non-uniform 
variation of the growth rates as a function of the drift velocity, showing that growth 
rates of WHF branch are conditioned by the thermal velocity of the resonant beaming 
electrons, satisfying $v_{res} \gtrsim u_b$. Moreover, increasing the drift velocity 
and plasma beta induces a transition regime of interplay of both WHF and firehose-like 
branches of HF instability.

\subsection{Quasilinear Analysis}\label{Sec:4.2}
Quasilinear (QL) analysis allows us to follow and understand temporal evolution of the enhanced fluctuations, 
i.e., the increase of the wave energy density $\delta B(t)/B_0^2$ up to the saturation, as well as the 
reaction of these fluctuations back on the VDFs of plasma particles, describing their relaxation
and, eventually, their thermalization. To do so, we solve the the set of QL equations (\ref{e9}) 
and (\ref{e10}) for three distinct sets of plasma parameters, as mentioned above as cases 1, 2 
and 3.
%
\subsubsection{Case 1}\label{sec:4.2.1}
%
\begin{figure*}
\centering
\includegraphics[scale=0.72, trim={2.9cm 6.2cm 2.5cm 6.8cm}, clip]{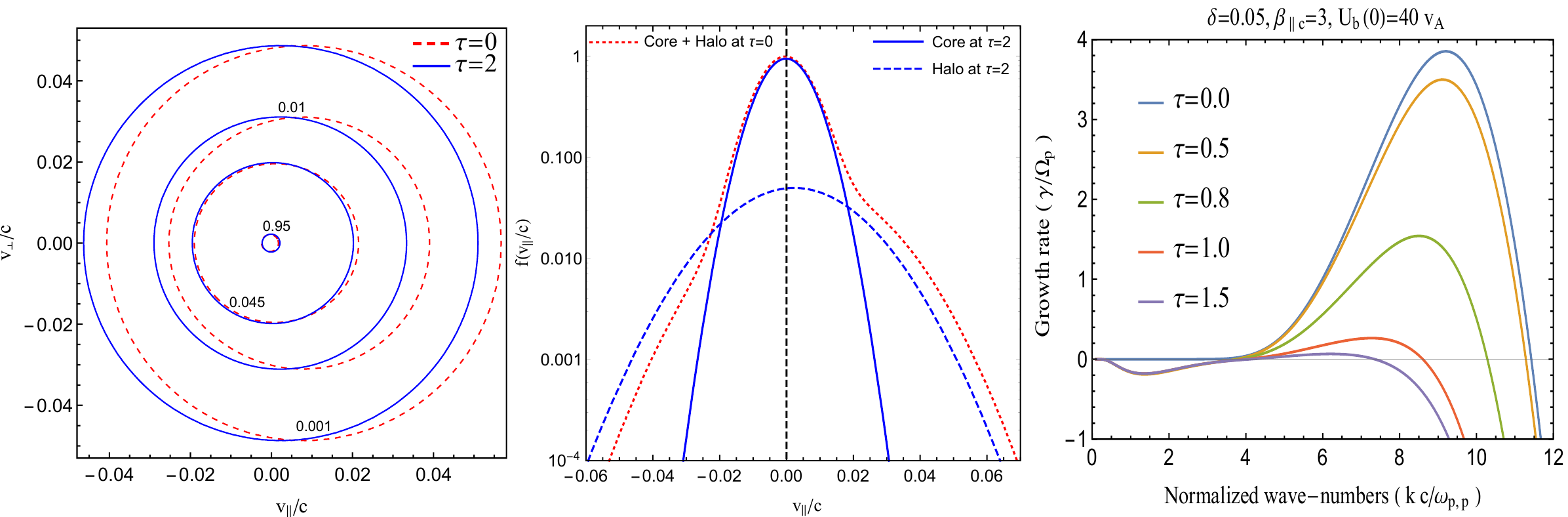}
\caption{Case~1: Contours (left) and parallel cuts (middle) of 
electron velocity distribution ($\beta_{c,\parallel}=1$), initially ($\tau=0$) 
and after $\tau=2$; WHF growth rates for $\beta_{c,\parallel}=3$ (right) at 
different times relevant for the evolutions in Figure \ref{f3}, right panels.}\label{f4}
\end{figure*}
\begin{figure}
\centering
\includegraphics[scale=0.45, trim={2.6cm 2.5cm 0cm 2.6cm}, clip]{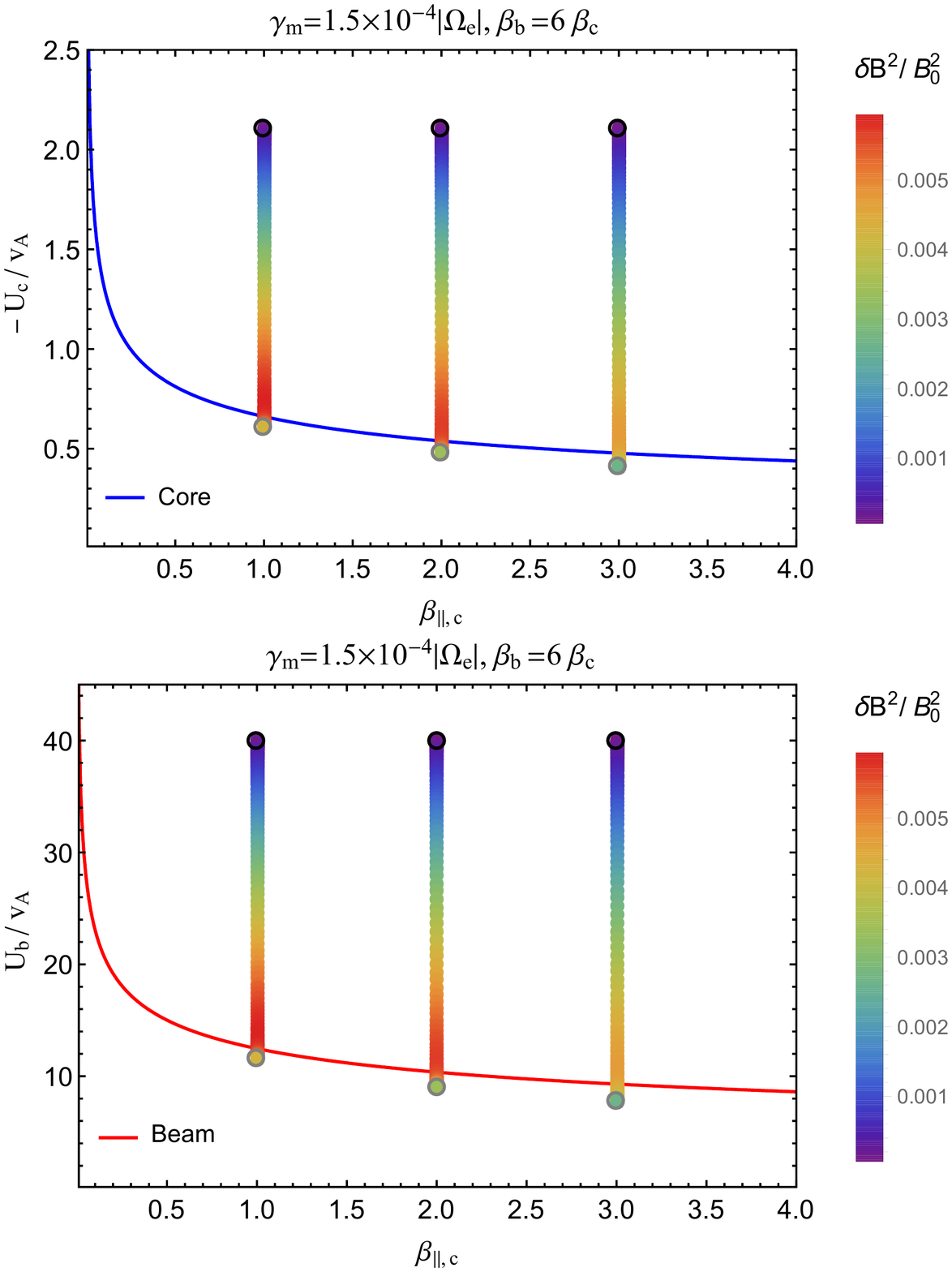}
\caption{Case 1: dynamical decreasing paths for the core and beam electrons in ($u_{a},~ \beta_{\parallel c}$)$-$space, whose initial states are shown with black circles. Final positions after saturation are indicated with gray circles, and the magnetic wave energy level is color-coded.} \label{f5}
\end{figure}
In case 1 we assume different initial conditions as given by different values of $\beta_{\parallel 
c}(0)$ (implying different $\beta_{\parallel b}=6~\beta_{\parallel c}$), and study the evolution of 
the fluctuating power and the main drivers of the instability, i.e., drift velocities of core 
and beaming electrons, by assuming that their temperature anisotropies do not change in time
and remain isotropic, i.e., $A_{c,b}(\tau)=1.0$. Thus, Figure \ref{f3} displays temporal evolution 
of the drift velocities for the core ($U_c$) and beam ($U_b$), the wave energy density ($\delta 
B(t)/B_0^2$), and the heat flux ($q_e/q_{max}$)  as functions of the normalized time $\tau= \Omega_p t$ 
and for different initial conditions: $\beta_c(0)=1$ (left), $2$ (middle), $3$ (right). 
Both components are relaxed, as their relative drifts are both reduced in time, up to the 
saturation of the instability, when the enhanced fluctuating power starts diminishing.
An increase of the initial $\beta_c(0)$ accelerates these mechanisms, but reduces the 
effective anisotropy \citep{Shaaban2018HFA} leading to lower drifts  and lower levels of fluctuations after saturation.
Bottom panels show the decrease of heat flux with similar time profiles.

Figure~\ref{f4} displays contours (left) and parallel cuts (middle) of the electron distributions 
at initial and final times, i.e.\ $\tau=0$ (red-dotted) and $2$ (blue-solid), respectively, for 
$\beta_{\parallel c}(0)=1$ corresponding to the results in Figure~\ref{f3}, left panels. For both 
snapshots contour levels represent $10^{-3}$, $10^{-2}$, $0.045$, and $0.95$ of $f_{e,max} = 1$. 
It is obvious that after the saturation of the WHF instability, e.g.\ at $\tau=2$, the drift 
velocities are both very low and the VDF is stable. This reduction of the relative drift velocities 
at later times ($\tau=2$) is more apparent in middle panel showing the parallel cuts of the electron
distributions. Moreover, right panel displays the WHF growth rates for $\beta_{\parallel c}(0)=3$ 
(corresponding to the results in Figure \ref{f3}, right panels), at intermediary time steps. As
expected, these growth rates decrease due to a decrease of the drift velocities in time.
The variation of the wave frequency is negligible (not shown here).

\begin{figure*}
\centering
\includegraphics[scale=1.05, trim={2.6cm 13.8cm 2cm 2.5cm}, clip]{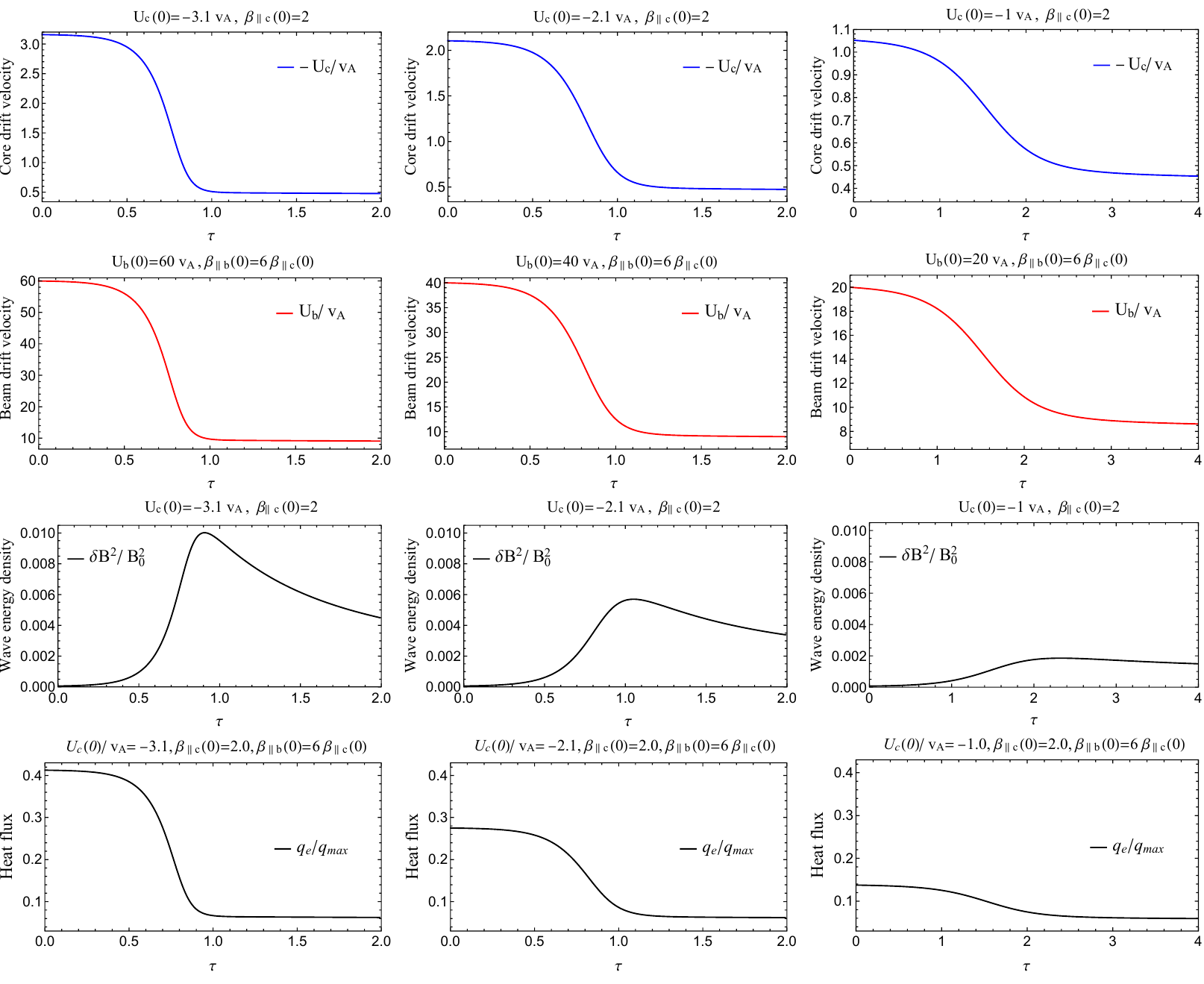}
\caption{Case~2: QL time evolution of the drift velocities for 
core ($U_c$) and beam electrons ($U_b$), the wave energy density ($\delta B(t)/B_0^2$), and the
heat flux ($q_e/q_{max}$) for different initial conditions: $U_c=-3.1~v_A$ (left), $U_c=-2.1~v_A$ 
(middle), $U_c=-1.0~v_A$ (right).}\label{f6}
\end{figure*}

Dynamical paths of the drift velocities $U_a/v_a$ as a function of $\beta_{\parallel c}$ are 
displayed in Figure \ref{f5}, for both the core ($"a = c"$, top panel), and beam ($"a = b"$, 
bottom panel). Black circles indicate initial positions, while gray circles mark 
final states after saturation. The variation of magnetic wave energy $\delta B^2/B_0^2$ is 
color coded. The final states of the dynamical paths end up very close to the instability thresholds, i.e., 
the lowest drift velocities predicted by the linear theory (unstable regimes are situated above the 
thresholds). These temporal profiles clearly show the relaxations of the core and beam drift 
velocities towards the most stable regime in agreement with the drift velocity thresholds 
predicted by the linear theory. These velocity thresholds are derived for a maximum growth rate 
$\gamma_{max}=0.27~\Omega_p\approx1.5\times10^{-4}|\Omega_e|$ , and are well fitted to \citep{Shaaban2018HFA}
\begin{equation}\label{e17}
-U_{c}/v_A=\frac{s}{\beta_{\parallel~c}^{~\alpha}},~~~\&~~ U_{b}/v_A=\frac{s}{\beta_{\parallel~ c}^{~\alpha}}.
\end{equation}
with ($s,~\alpha$)~= ~($0.66,~0.30$) for the core ($a=c$) and  ($s,~\alpha$) = ($12.47,~0.27$)  for the beam 
($a=b$) components.

\subsubsection{Case 2}\label{Sec:4.2.2}
%
\begin{figure*}
\centering
\includegraphics[scale=0.72, trim={2.9cm 6.2cm 2.5cm 6.8cm}, clip]{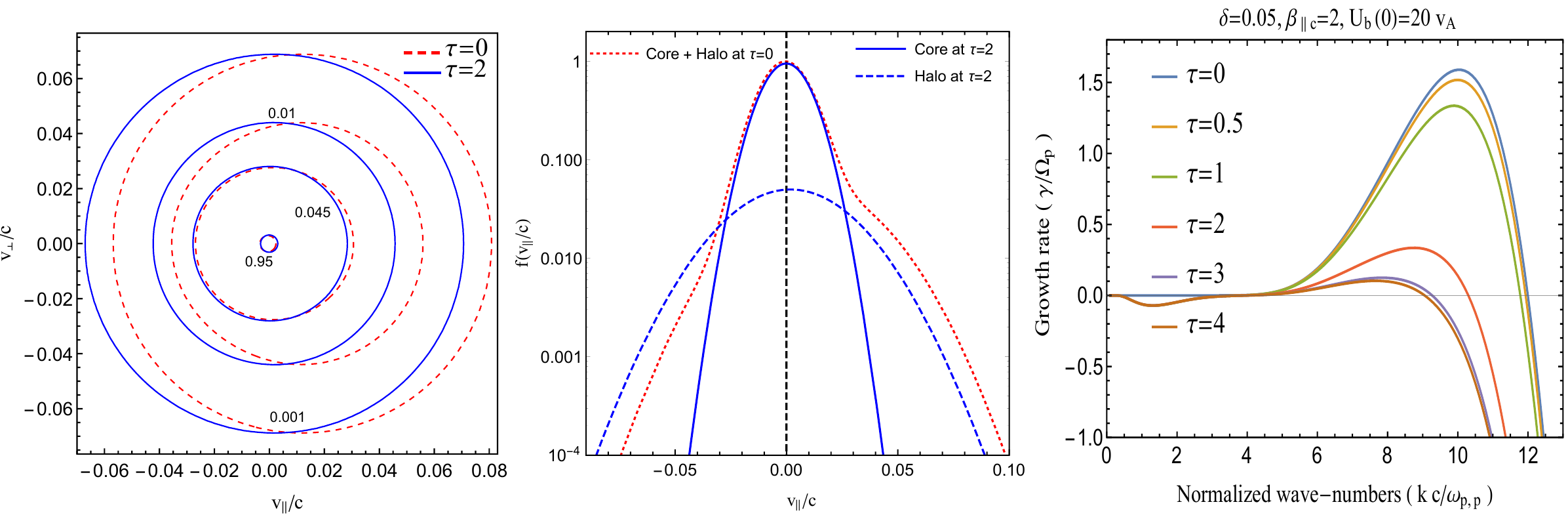}
\caption{Case 2: Contours (left) and parallel cuts (middle) of 
electron velocity distribution ($U_{c}=-3.1~v_A$), initially ($\tau=0$) 
and at $\tau=2$; WHF growth rates for $U_c=-1.0~v_A$ (right) at different times 
relevant for the evolutions in Figure \ref{f6}, right panels.}\label{f7}
\end{figure*}
Here our QL analysis starts from different conditions, assuming initially three different core 
drift velocities $U_c/v_A=-3.1,~-2.1,~-1$ (implying different beaming velocity $U_b/v_A=60,~40,~20$). 
Figure~\ref{f6} shows time evolution for the drift velocities of the core ($U_c$) and beam ($U_b$), 
the corresponding time variation of the wave energy density ($\delta B^2/B_0^2$) and the normalized 
heat flux ($q_e/q_{max}$). We assume again that temperature anisotropy is not affected by the growing
fluctuations. After the saturation the core and beam drift velocities end up to almost the same 
velocities, regardless the value of the the initial drift velocity. However, lower drift velocities, 
i.e. $U_c=-1.0~v_A$ and $U_b=20~v_A$, need longer time to relax. The associated wave energy density 
and the normalized heat flux increase with increasing the core drift velocity, confirming the 
enhancement of the WHF growth rates in Figure~\ref{f2}. 

\begin{figure}
\centering
\includegraphics[scale=0.45, trim={2.5cm 4cm 1cm 2.5cm}, clip]{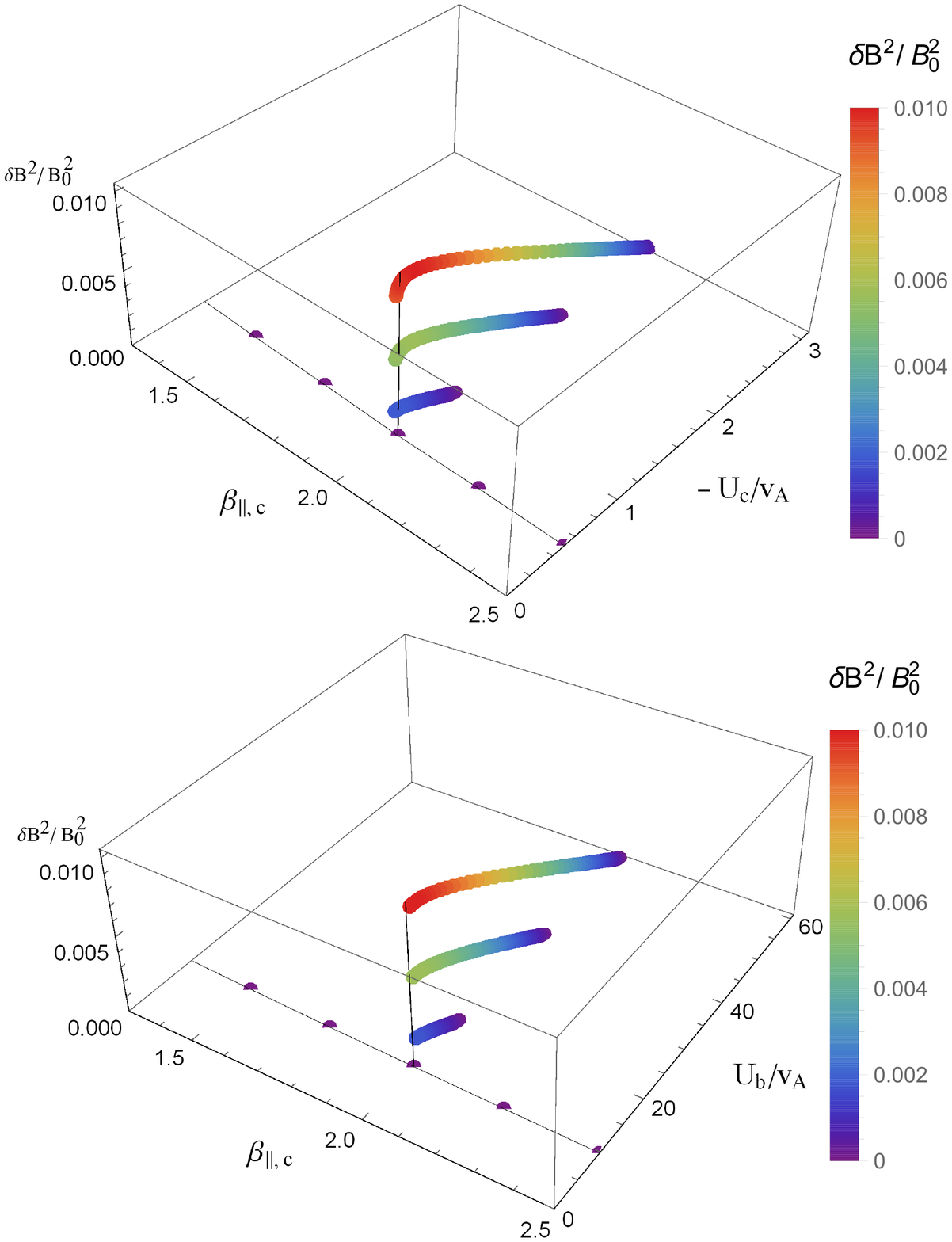}
\caption{Case 2: 3D display dynamical paths of the drift velocities for the core (top) and beam (bottom), and the varying level of magnetic wave energy with colors . 
Initial states are shown with black circles, while final positions after saturation are indicated with gray circles which follow the profile of the instability thresholds at 
$\beta_{c,\parallel}=2$.}
\label{f8}
\end{figure}

In Figure~\ref{f7}, contours (left panel) and parallel cuts (middle panel) represent the 
initial and final states (i.e.\ $\tau=0$ and $\tau=2$, respectively) of the eVDFs in the QL 
evolution of WHF instability for $U_c (0)=-3.1~v_A$, see also Figure~\ref{f6}. Initial drift 
velocities of the core and beam are regulated by the enhanced fluctuations, and distribution 
becomes less anisotropic and, therefore, more stable at $\tau=2$ (blue solid contours). 
Right panel displays the WHF growth rates for $U_c=-1.0~v_A$, for different times up to the 
saturation and after, corresponding to the results in Figure~\ref{f6}, left panels. 
These growth rates are decreasing in time coresponding to a decrease of the core and 
beam drift velocities, see Figure \ref{f6}. The time variation of the wave frequency is 
negligible and is not shown here.

In case 2, comparison of the instability thresholds (predicted by the linear theory) and dynamical 
paths of the core and beam drift velocities from QL analysis is made for the same value of the 
core plasma beta, i.e.\ $\beta_{\parallel c}=2$, and is therefore less straightforward. 
Figure~\ref{f8} displays 3D dynamical paths of the core and beam drift velocities in order 
to avoid their overlap. Threshold conditions are indicated by dotted lines in the horizontal 
plane $U_a/v_A - \beta_{c,\parallel} $, and final states align to the vertical line of drift 
velocities corresponding to maximum wave energy densities. Dynamical paths of the drift 
velocity for both core (top) and beam (bottom) end up very close to the instability threshold 
predicted by the linear theory at the same condition of saturation for the wave energy density. 

\begin{figure*}
\centering
\includegraphics[scale=1.1, trim={2.7cm 20.7cm 2.2cm 2.5cm}, clip]{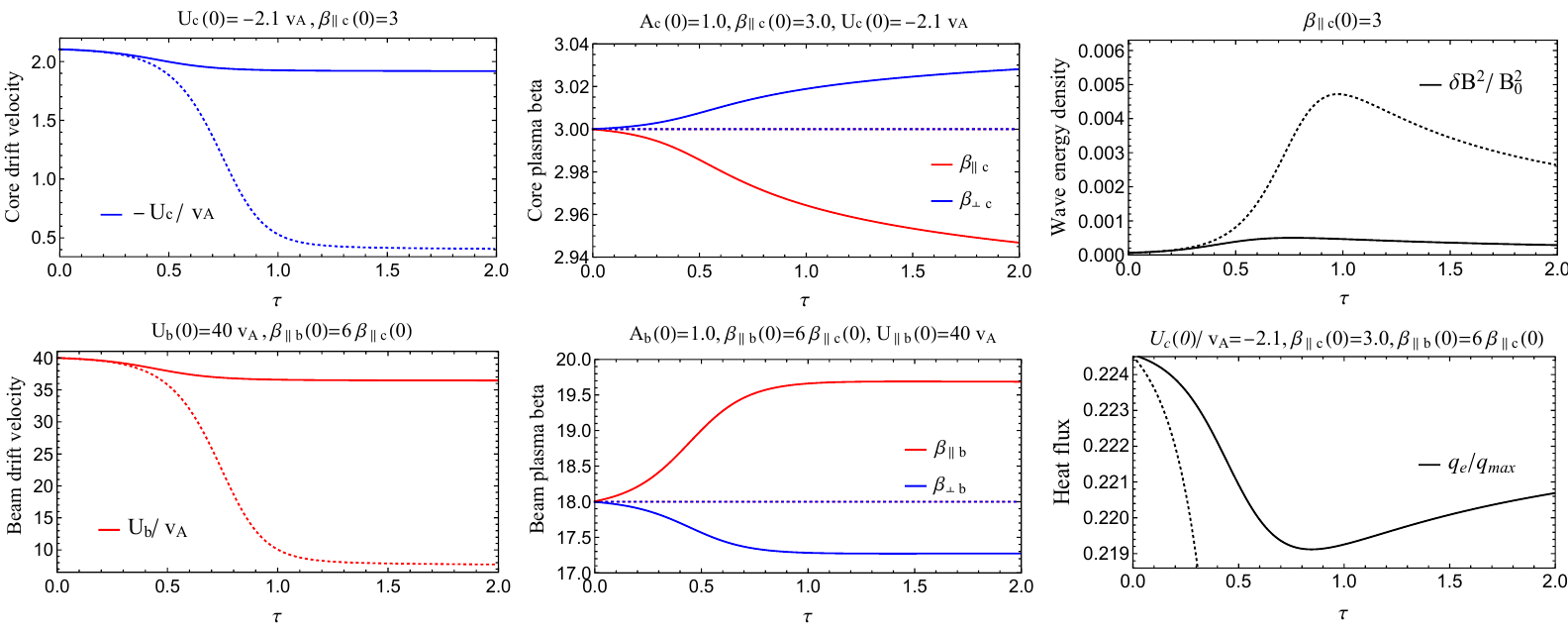}
\caption{Case 3: QL time evolution of the drift velocities (left), 
plasma betas (middle), the wave energy density (top-right) and the heat flux (bottom-right) for $U_c=-2.1~v_A$ and $\beta_c=3.0$. }
\label{f9}
\end{figure*}
\begin{figure*}
\centering
\includegraphics[scale=0.71, trim={2.9cm 6.2cm 2.6cm 6.9cm}, clip]{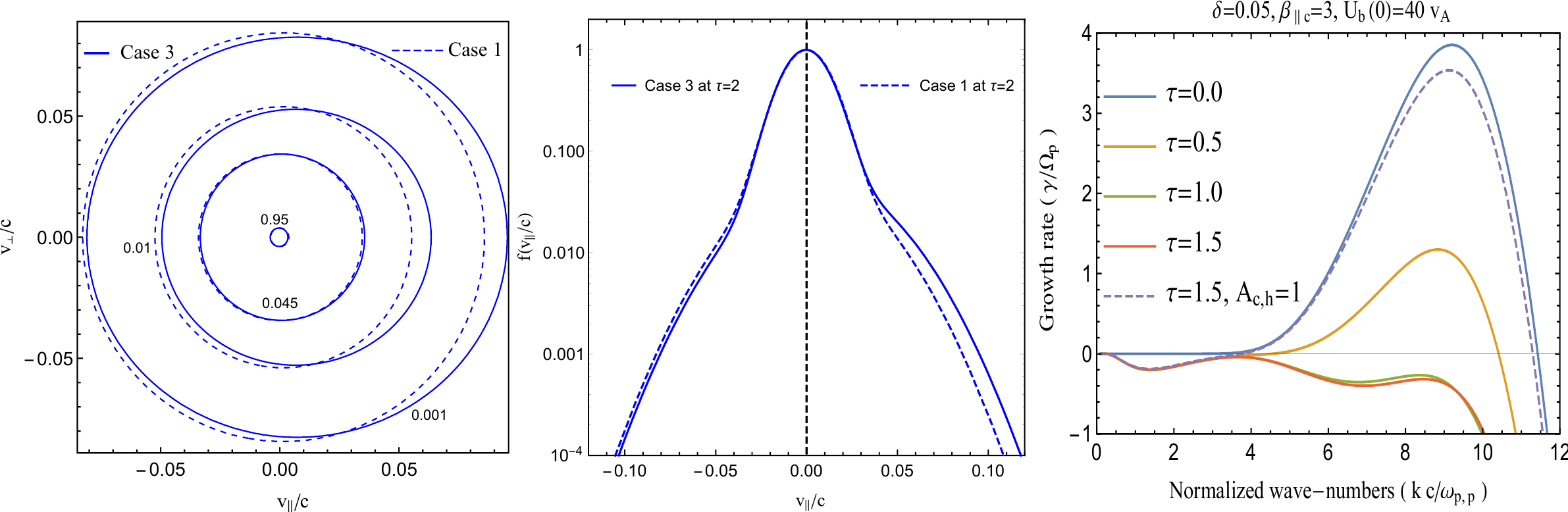}
\caption{Case 3: Contours (left) and parallel cuts (middle) of 
electron velocity distribution, initially ($\tau=0$) and at $\tau=2$; WHF growth rates for $U_c=-1.0~v_A$ (right) at different times relevant for the evolutions in Figure \ref{f9}.}\label{f10}
\end{figure*}
%

%
\subsubsection{Case 3}\label{Sec:4.2.3}
%
%
\begin{figure}
\centering
\includegraphics[scale=0.45, trim={2.6cm 2.5cm 0cm 2.6cm}, clip]{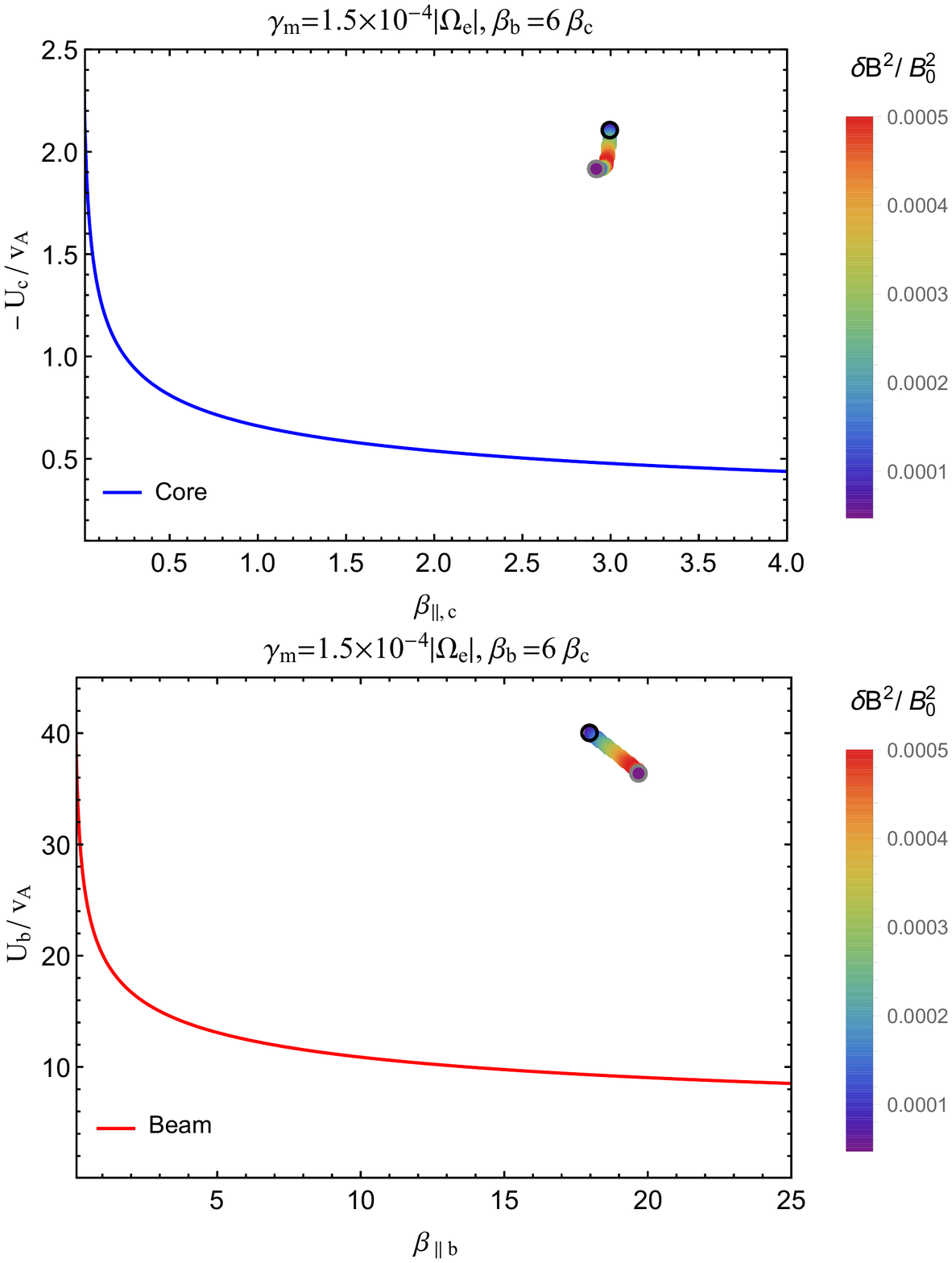}
\caption{The same as Figure~\ref{f5}, but for case 3.} \label{f11}
\end{figure}

Suppose now, more realistically, that all plasma parameters, e.g., drift velocities ($u_{c,b}$), 
plasma beta parameters ($\beta_{c,b}$), may vary in time (i.e., normalized time $\tau=\Omega_p t$). 
Figure ~\ref{f9} presents, by comparison, time evolutions for the core and beam drift velocities 
(left panels), the plasma betas (middle panels), and the corresponding wave energy density (top-right) 
and heat flux (bottom-right) for an initial $\beta_{\parallel c} (0)=3.0$. Comparison includes 
cases 1 (dotted lines) and 3 (solid lines). Allowing electron temperatures and, implicitly, 
plasma betas to vary in time markedly reduces the relaxation of the core and beams drift velocities
(solid lines) to values only slightly below the initial conditions but much higher than those 
obtained after relaxation in case 1 (dotted lines). However in this case temperatures do not remain
isotropic, but change under the effect of enhanced WHF fluctuations leading to small deviations from 
isotropy. The core and beam components reach opposite temperature anisotropies after relaxation, as 
indicated by their parallel (red) and perpendicular (blue) plasma betas in middle panels. The core 
shows an excess of perpendicular temperature $A_c(\tau=2)=\beta_{\perp c}/\beta_{\parallel c}(\tau=2)
=1.024$, while the beam shows an excess of parallel temperature $A_b(\tau=2)=\beta_{\perp b}/\beta_{\parallel 
b}=0.877$. \cite{Shaaban2018HFA} have indeed shown that WHF instability is markedly inhibited by such a
temperature anisotropy of the beam, that may explain, comparing to case 1, the saturation at lower wave 
energy densities, a modest relaxation at higher drift velocities and a minor reduction of the heat-flux. 
The opposite anisotropies gained by the core and beam components can be 
explained by a series of selective mechanisms: (a) heating of the core electrons in perpendicular 
direction by resonant interactions with whistlers, (b) scattering and diffusion in velocity space
of the beam electrons leading to a plateau formation and an effective temperature anisotropy 
$T_{b,\parallel} > T_{b,\perp}$.

In Figure~\ref{f10}, we display, for comparison, contours and parallel cuts of the eVDFs in the 
later stages after saturation, (i.e., $\tau=2$) for case 1 (dotted lines) with $\beta_{c 
\parallel}=3$ and case 3 (solid lines). These final states are markedly different, in case 3 
we can still observe an anisotropic distribution drifting in parallel direction, while 
in case 1 it is more isotropic (and therefore more stable). Right-panel in Figure~\ref{f10} shows the WHF 
growth rates for case 3 at different intermediary times. These growth rates decrease in time, 
and the WHF mode is damped at $\tau=1.5$. In order to understand the role played by the induced 
temperature anisotropies, e.g., in Figure~\ref{f9} middle panels, with dashed line we add the WHF 
growth rate obtained for the same drifting velocities of the core and beam at $\tau=1.5$ but when 
temperatures of these populations are maintained constant and isotropic (similar to cases 1 and 2). 
This growth rate display a considerable peak, while for case 3 when the core exhibit $T_{c,\perp} >
T_{c,\parallel}$ and the beam $T_{b,\perp} < T_{b,\parallel}$, WHF mode is damped. 
Allowing temperatures and implicitly plasma betas to vary in time inhibits the growth rates 
and leads to a faster saturation of the instability. Without artificial constraints to keep constant 
particle temperatures (or plasma beta parameters), the QL results seem to be more 
realistic than those obtained in cases 1 and 2, and suggest a new self-inhibiting effect of 
WHF instability. This saturation results from multiple effects combining a minor relaxation of 
drift velocities with a transverse heating of the (highly dense) core, and an additional (elastic) 
scattering of the beam \citep{Marsch2006}, which together contribute to a reduction of the 
effective anisotropy and inhibit the instability.  

Opposite variations of parallel plasma beta parameters, towards lower values for the core (top) 
and higher values for the beam (bottom) are also shown in Figure~\ref{f11} by the dynamical paths 
of the drift velocities for the core (top) and beam (bottom). However, in this case final states 
do not approach the instability thresholds predicted by the linear theory, which have no relevance 
in this case. This result 
demonstrates the importance of an extended QL approach of WHF instability, which can 
realistically unveil long-term evolution of the enhanced fluctuations with multiple effects on 
the velocity distributions of electrons (i.e., thermalization, cyclotron heating). Linear theory 
can predict only a reduction of the drift velocities towards lowest threshold values, but a QL 
approach may quantify additional transfers of energy between wave fluctuations and particles, 
as shown by the time evolutions of the higher order moments of the velocity distribution, 
e.g., temperatures and heat-flux.

%
\section{Discussions and Conclusions}\label{Sec:5}
%
In the present paper we have characterized the quasilinear evolution of the whistler heat-flux instability 
driven by two counter-beaming Maxwellian electron populations, resembling the velocity distributions observed 
in the space plasmas. Central component is the highly dense core but the additional beaming or strahl population
is mainly responsible for the electron heat-flux in the solar wind. The whistler heat-flux instability
is selfgenerated when the relative beaming or drift velocity does not exceed thermal speed of 
beaming electrons. However, the main interest is to understand if growing fluctuations act back on 
the electron beams, reducing their anisotropy and regulating the heat-flux. Our present results from an 
extensive quasilinear study of the whistler heat-flux instability may offer valuable answers to these 
questions. 

Section 4 presents the results of a parametric study on the influence of the initial conditions, 
i.e., macroscopic plasma parameters like drifting velocities and plasma beta parameters for the 
core and beam, on the saturation of this instability and the relaxation of the electron distribution. 
In the firsts two cases we have allowed only for time variations of the drift velocities (the 
source of free energy), while temperatures are kept constant. Higher plasma betas assumed initially 
for the core electrons (case 1) seem to stimulate the instability maximum growth rate in the 
linear phase, but restrain to range of unstable wave-numbers. Moreover, growing fluctuations 
become less robust saturating faster and at lower fluctuating field energy densities. The 
corresponding relaxations of counter-beams and the heat-flux are also faster leading to lower 
levels. In fact, particle thermalization is indeed expected to suppress beaming instabilities, 
but for the whistler heat-flux instability this effect becomes obvious only in the long term 
evolution. The influence of drifting velocity (case 2) is less controversial, e.g., higher values 
of $U_c$ enhance not only the growth-rates but markedly stimulate growing fluctuations to high 
energy densities, and implicitly, accelerates the inhibition of the drifts and the electron 
heat-flux. In both these two cases the saturated stage is described by more stationary 
distributions, which carry lower heat-flux, and approach, as expected, the drift velocity 
thresholds predicted by linear theory. 

Case 3 is more realistic as we have enabled for time variations of all plasma parameters, i.e., 
temperatures and drift velocities, as may result from the interactions of electrons with 
growing fluctuations. The relaxation of the core and beams drift velocities is very modest 
in this case, to values only slightly below the initial conditions. However, temperatures also 
change leading to small deviations from isotropy, and opposite anisotropies of the core and 
beam components, which actually explains the faster inhibition of the instability \citep{Shaaban2018HFA} 
and minor relaxation of the beam and the heat-flux. Triggered by the whistler fluctuations, 
these effects result from the concurrence of the core electrons heating by resonant interactions 
with whistlers, and elastic scattering of the beam electrons and their diffusion in velocity space 
leading to a plateau formation and an effective temperature anisotropy $T_{b,\parallel} > T_{b,\perp}$. 

We can state that our QL approach in case 3 enables a number of clear conclusions.
Linear theory cannot capture the energy transfer between the electron populations, 
and consequently cannot describe correctly the saturation of whistler heat flux 
instability and the relaxation of velocity distribution. It is QL theory that offers a 
more complete and realistic picture, showing that this relaxation is determined
by a concurrent effect of the induced temperature anisotropies, which are small but 
efficient, and an additional minor relaxation of relative drift velocities. Whistler 
heat-flux instability is very sensitive to the initial conditions (i.e., with 
two lower and upper thresholds of relative drift, see \cite{Gary1985, Shaaban2018HF, Shaaban2018HFA}, and 
the resulting wave fluctuations have only modest or very weak intensities, and, 
implicitly, minor effects on particle distributions. 
In our present analysis, initial temperatures of counter-beaming electron populations are assumed 
isotropic, an assumption supported by recent observations, e.g., in \cite{Tong2019}, which 
identify whistler fluctuations in association with core-beam electron distributions and only  
beams with isotropic temperatures. It is also claimed that a small anisotropy $A_b \lesssim 1 $ may 
prevent the whistler heat-flux instability even in the presence of a considerable drift velocity of 
the core, which can also be explained by the quasi-stable states after relaxation (case 3).
To conclude, these observations are in agreement with our results in case 3, which (i) show that 
the saturation of the whistler heat-flux may occur via the induced temperature anisotropies 
for the core and beam, and (ii) confirm a minor implication of this instability in the regulation 
of electron heat-fluxes, as suggested by recent studies \citep{Horaites2018, Vasko2019} indicating 
instabilities of oblique modes, e.g., kinetic Alfv\'en and magnetosonic waves, as potentially more 
efficient in scattering and suppressing the heat-flux of electron strahl.
 
\section*{Acknowledgements}

The authors acknowledge support from the Katholieke Universiteit Leuven, Ruhr-University Bochum, and 
Alexander von Humboldt Foundation. These results were obtained in the framework of the projects 
SCHL 201/35-1 (DFG-German Research Foundation), GOA/2015-014 (KU Leuven), G0A2316N (FWO-Vlaanderen), 
and C 90347 (ESA Prodex 9). S.M. Shaaban would like to acknowledge the support by a Postdoctoral 
Fellowship (Grant No. 12Z6218N) of the Research Foundation Flanders (FWO-Belgium). P.H.Y. acknowledges the BK21 Plus grant (from NRF, Korea) to Kyung Hee University, and financial support from GFT Charity Inc., to the University of Maryland.



\bibliographystyle{mnras}
\bibliography{papers} 


\bsp	
\label{lastpage}
\end{document}